\documentclass[12pt,preprint]{aastex}
\usepackage{graphicx}
\usepackage{epsfig}
\usepackage{rotating}

\def\Msol{\thinspace\hbox{$\hbox{M}_{\odot}$}}

\def\a4{\hsize 17.0cm \vsize 25.cm}

\begin{document}

\title{On the origin of the neutral hydrogen supershells: the
       ionized progenitors and the limitations of the multiple
       supernovae hypothesis}

\author{
Sergiy Silich \affil{Instituto Nacional de Astrof\'\i sica Optica y
Electr\'onica, AP 51, 72000 Puebla, M\'exico; silich@inaoep.mx}
Federico Elias \affil{Instituto de Astronom\'\i a. Universidad
Nacional Aut\'onoma de M\'exico. AP 70 - 264, 04510 M\'exico D.F.;
felias@astroscu.unam.mx} \and Jos\'e Franco \affil{Instituto de
Astronom\'\i a. Universidad Nacional Aut\'onoma de M\'exico. AP 70 -
264, 04510 M\'exico D.F.; pepe@astroscu.unam.mx}}

\begin{abstract} Here we address the question whether the ionized shells
associated with giant HII regions can be progenitors of the larger
HI shell-like objects found in the Milky Way and other spiral and
dwarf irregular galaxies. We use for our analysis a sample of 12 HII
shells presented recently by Rela\~no et al. (2005, 2007). We
calculate the evolutionary tracks that these shells would have if
their expansion is driven by multiple supernovae explosions from the
parental stellar clusters. We find, contrary to Rela\~no et al.
(2007), that the evolutionary tracks of their sample HII shells are
inconsistent with the observed parameters of the largest and most
massive neutral hydrogen supershells. We conclude that HII shells
found inside giant HII regions may represent the progenitors of
small or intermediate HI shells, however they cannot evolve into the
largest HI objects unless, aside from the multiple supernovae
explosions, an additional energy source contributes to their
expansion.
\end{abstract}


\section{Introduction}

The origin of numerous holes and shells detected in the distribution
of neutral hydrogen in spiral and dwarf galaxies (Heiles, 1980;
Brinks \& Bajaja, 1986; Puche et al., 1992) is a long standing
problem (Heiles, 1984; Tenorio-Tagle \& Bodenheimer, 1988). Heiles
(1984), Brinks \& Bajaja (1986), Puche et al. (1992), Mashchenko et
al. (1999), Ehlerov\'a \& Palou\v s (2005) detected in the Milky
Way, M31 and in the dwarf irregular galaxy Holmberg II (HoII)
hundreds of neutral hydrogen shells whose radii range from a few
tens to a thousand parsecs. On the other hand, Fabry-Perot
observations of the ionized gas kinematics in many galaxies revealed
a number of shells whose radii reach a few hundred parsecs and whose
velocity pattern suggests an expansion with velocities up to 60 --
70~km $s^{-1}$ (see, for example, Lozinskaya 1992; Chu et al. 1990;
Oey \& Massey, 1995; Valdez-Guti\'errez et al. 2001; Naz\'e et al.
2001; Lozinskaya et al. 2003;  Rela\~no \& Beckman, 2005 and
references therein). The majority of these shells is associated with
interior stellar clusters. The standard model (McCray \& Kafatos,
1987; Mac Low \& McCray 1988) was constructed with the energy
injection from multiple supernovae explosions and stellar winds
occurring in young stellar clusters which are often found inside
small and intermediate sized shells. In such a case the energy
supplied by supernovae and individual stellar winds is thermalized
inside the parental stellar cluster, resulting in a high central
overpressure which drives a high velocity outflow (the star cluster
wind). This outflow, when interacting with the ambient interstellar
medium (ISM), forms a leading and a reverse shocks which are
separated by a contact discontinuity (see Figure 1). The
interstellar gas collected by the outer shock forms an expanding
shell which moves because the thermal pressure in the region C,
between the reverse shock and the contact discontinuity, exceeds the
ISM pressure. The swept-up gas cools rapidly and forms an expanding
shell photoionized by the Lyman continuum from the embedded cluster.
The number of ionizing photons rapidly drops with the star cluster
age (Leitherer et al. 1999) and eventually (after $\sim 10$~Myrs)
the driving cluster (or clusters) will find itself embedded inside a
slowly expanding neutral hydrogen shell. The shell radius and
velocity depend on the amount of energy released by the cluster and
on the parameters of the ambient interstellar medium (see review by
Bisnovatyi-Kogan \& Silich, 1995 and references therein).

While the standard model is broadly consistent with the parameters
of many small and intermediate sized structures found around young
stellar clusters and OB associations (see, for example, the
discussion of the superbubble growth discrepancy in Oey \& Garc\'\i
a-Segura, 2004 and the analysis of the XMM-Newton observations of
the 30 Dor C in Smith \& Wang 2004), it meets a profound energy
problem when applied to larger structures whose radii are comparable
to the characteristic Z-scale of the density distribution in the
host galaxy. For instance, Rhode et al. (1999) found that, in the
case of the HoII galaxy, the stellar clusters found inside HI holes
are unable to remove gas from these regions. Kim et al. (1999) found
only a weak correlation between the Large Magellanic Cloud (LMC)
neutral hydrogen holes and the HII regions and concluded that the
hypothesis of multiple supernovae is inconsistent with their data.
Hatzidimitriou et al. (2005) cross-correlated the positions of 509
young neutral hydrogen shells detected in the Small Magellanic Cloud
(SMC) with the locations of known OB-associations, Wolf-Rayet stars,
supernova remnants and stellar clusters. They found that 59 shells
have no young stellar objects associated with them despite the
distributions of their radii and expansion velocities are consistent
with those predicted by the multiple SNe model. They concluded that
turbulence may be a promising mechanism that would allow us to
understand the origin of these objects, but a quantitative
comparison of the existing theory with observations is not possible
at this moment.

Crosthwaite et al. (2000) examined the distribution of HI in the
nearly face-on Scd galaxy IC 342 and did not find either the
kinematic signatures specific to the expanding shells, or
indications on the distortion of the observed structures by the
differential galactic rotation, which is expected to be noticeable
if the observed structures expand because of the energy output
provided by the embedded stellar population (Palous et al. 1990;
Silich, 1992). Therefore Crosthwaite et al. (2000) have concluded
that the observed flocculent structure is formed via gravitational
instabilities in a turbulent galactic disk, as suggested by Wada \&
Norman (1999), and not because of the energy output provided by the
stellar component of the galaxy. Silich et al. (2006) have compared
parameters of a $\sim 500$~pc radius HI ring detected in dwarf
irregular galaxy IC 1613 with a combined energy output provided by a
number of OB associations found inside the structure. They did not
find a noticeable expansion velocity and concluded that the observed
radius and mass of the structure are inconsistent with the SNe
hypothesis.

Several other mechanisms that allow to shape the ISM of the host
galaxy into large shell-like structures similar to those observed in
nearby spiral and irregular galaxies, and that do not require
violent stellar activity, include collisions of high velocity clouds
with galactic disks (Tenorio-Tagle, 1981; Comeron \& Torra, 1992),
non-linear instabilities in the self-gravitating turbulent galactic
disks (Wada et al. 2000; Dib \& Burkert, 2004) and more exotic
mechanisms such as the distortion of the ISM by powerful gamma-ray
bursts (Efremov et al. 1999).

On the other hand, several modifications of the multiple SNe
hypothesis have been suggested by different authors, in an attempt
to reconcile the observed parameters of large HI structures with
those predicted by the multiple supernovae model. Elmegreen \&
Chiang (1982) added the effects of the radiation pressure from field
stars, that is able to provide additional expansion. Palous et al.
(1990) added the effects of galactic rotation that distorts and
stretches the shell along the direction of rotation.
McClure-Griffiths et al. (2002) found that some HI shells are
located between the spiral arms of the Galaxy. They suggested then
that density gradients that occur between the spiral arms and the
interarm medium could result in the enhancement of the predicted
sizes and migration of large shells from spiral arms into the
interarm medium.

More recently, Rela\~no et al. (2007) have suggested that the
population of ionized H$_{\alpha}$ shells associated with large HII
regions (Rozas et al. 1996; Rela\~no \& Beckman, 2005) may represent
the precursors of the larger HI structures, and claimed that a very
simplified analytic model is enough to explain larger HI objects.
They assumed that young shells observed in H$_{\alpha}$ emission
evolve in an energy dominated regime until the beginning of the
supernova explosion phase, then make a transition to a momentum
dominated stage and continue to expand with the momentum supplied by
the embedded cluster during the initial time. The idea is not new
(Dyson, 1980; Bruhweiler et al. 1980) and was used by Gil de Paz et
al. (2002) and Silich et al. (2002), who discussed the origin of a
giant ($R \approx 700$~pc) ring in the low-metallicity BCD galaxy
Mrk 86. However, the key problem with this idea is that the momentum
dominated regime is inefficient in driving bubble expansion because
the thermal pressure in zone C (Figure 1) is approximately equal to
the ram pressure of the ejected gas at the location of the reverse
shock. Given that ram pressure drops as $r^{-2}$, the driving
pressure equals the external pressure at a certain radius and the
expansion quickly stalls after that point. Therefore we wonder how
the early transition to the momentum dominated stage can solve the
``energy crisis'' that has been discussed by many authors during the
last decades.

Here we re-analyze the hydrodynamical model presented in Rela\~no et
al. (2007) keeping their assumption of a fast transition to the
momentum dominated stage. We show that the over-simplified
hydrodynamic equations and the ``statistical'' initial conditions
adopted by them, which are not consistent with parameters for the
largest HI structures, lead to overestimated expansion velocities
and radii of the shells. Our results do not support their contention
that H$_{\alpha}$ shells associated with the HII regions can be
progenitors of the largest HI structures in the absence of any
additional driving mechanism (for instance, radiation pressure from
field stars, as suggested by Elmegreen \& Chiang 1982).

The paper is organized as follows: in Section 2 we establish the
main equations of our hydrodynamical model, and compare them with
those used by Rela\~no et al. (2007). In Section 3 we first
determine the initial conditions required to perform the numerical
integration of the equations, and then present the resulting
evolutionary tracks, our main result that small shells found in
giant HII regions cannot be the progenitors of the largest neutral
hydrogen supershells detected in gaseous galaxies.

\section{Main equations}

The kinetic energy supplied by supernovae and stellar winds from
stellar clusters is thermalized by a shock, and this results in the
four-zone structure discussed in Figure 1 (Castor et al. 1975;
Weaver et al. 1977; Mac Low \& McCray 1988; Bisnovatyi-Kogan \&
Silich 1995 and references therein). If thermal pressure in zone C
suddenly drops, the radius of the reverse shock approaches the
contact discontinuity, and the expansion of the outer shell is then
supported directly by the momentum deposited by the stellar cluster
(see, for example, Koo \& McKee, 1992). Rela\~no et al. (2007)
suggested that interstellar shells expanding around star forming
regions reach this stage after a short while and then evolve in the
momentum-dominated mode. We follow their assumption in the next
sections, but note that it is in bad agreement with estimates of the
characteristic cooling time in zone C (e.g., Figure 1 presented in
Mac Low \& McCray 1988). In the 2D or 3D cases, this regime occurs
at the waist of the shell if a driving cluster is embedded into the
disk-like density distribution (Gil de Paz et al., 2002; Silich et
al., 2006). The expansion of the shell (or of some segments of the
shell) is then defined by mass and momentum conservation:
\begin{eqnarray}
      \label{eq1.a}
      & & \hspace{-1.0cm}
M = M_0 + \frac{4 \pi}{3} (R^3 - R^3_0) \rho_{ISM} + {\dot M}_{SC} (t - t_0) ,
      \\[0.2cm]
      \label{eq1.b}
      & & \hspace{-1.0cm}
\frac{\rm d}{{\rm d} t} (M u) = - 4 \pi R^2 (P_{ISM} - \rho_w V^2_{\infty}) ,
\end{eqnarray}
where $M$, $u$ and $R$ are the mass, expansion velocity and radius
of the shell, respectively. $R_0$ is the initial radius of the
shell, $t_0$ is the initial time and $M_0$ is the initial mass of
the shell. The second term is the mass of the interstellar gas swept
up by the expanding shell, and the last term in equation
(\ref{eq1.a}) is the amount of ejected matter that sticks to the
shell from the inside. The right-hand part of equation (\ref{eq1.b})
represents the difference between the ambient gas pressure,
$P_{ISM}$, and the driving ram pressure of the ejecta. ${\dot
M}_{SC}$ is the mass deposition rate provided by SNe explosions and
stellar winds, $\rho_w(R)$ is the density of the ejected matter when
it reaches the shell, and $V_{\infty}$ is the terminal speed of the
ejected matter.

We assume that the parameters of the cluster remain constant during
the evolution, and that the expansion velocity of the shell is much
smaller than that of the ejected matter, $u \ll V_{\infty}$. The
mass deposition rate, ${\dot M}_{SC}$, and the density of the
ejecta, $\rho_w(R)$, then are:
\begin{eqnarray}
      \label{eq2.a}
      & & \hspace{-1.0cm}
{\dot M}_{SC} = 2 L_{SC} / V^2_{\infty}
      \\[0.2cm]
      \label{eq2.b}
      & & \hspace{-1.0cm}
\rho_w = {\dot M}_{SC} / 4 \pi R^2 V_{\infty} ,
\end{eqnarray}
where $L_{SC}$ is the rate of mechanical energy supplied by supernovae
and stellar winds.
\begin{figure}[htbp]
\plotone{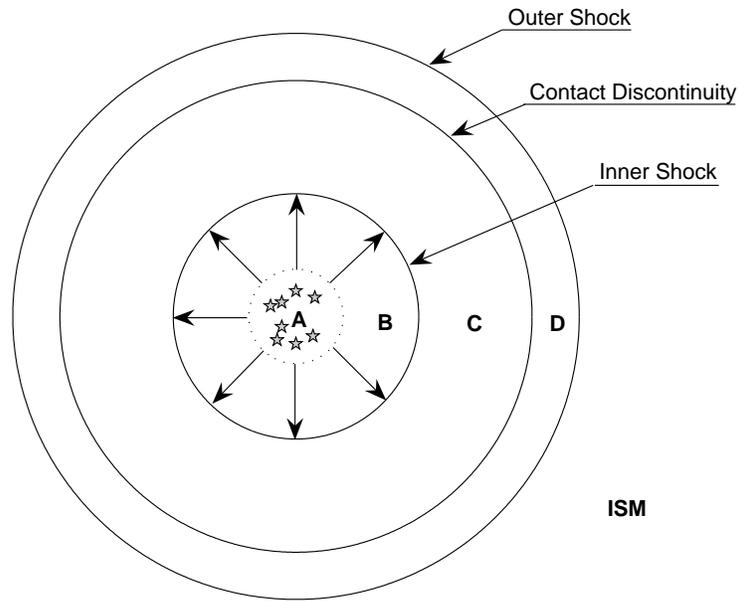} \caption{Schematic representation of the structure
formed in the ISM by multiple stellar winds and supernovae
explosions. The central zone (A) represents the stellar cluster
where stellar winds and supernovae release their energy. The
remaining concentric zones are the free-wind region (zone B), the
shocked wind region (zone C), the shell of swept up interstellar
matter (zone D) and the ambient ISM. When the shocked wind cools
rapidly, zone C vanishes and the outer shell is pushed away by the
momentum of the ejected matter.} \label{fig1}
\end{figure}

Combining equations (\ref{eq1.a}) and (\ref{eq1.b}) one obtains:
\begin{eqnarray}
      \label{eq3.a}
      & & \hspace{-1.0cm}
\frac{{\rm d}u}{{\rm d} t} = - \frac{4 \pi R^2 V^2_{\infty}
     (P_{ISM} + \rho_{ISM} u^2) - 2 L_{SC} (V_{\infty} - u)}
     {[M_0 + 4 \pi \rho_{ISM} (R^3 - R^3_0)/3] V^2_{\infty} +
      2 L_{SC} (t - t_0)}
      \\[0.2cm]
      \label{eq3.b}
      & & \hspace{-1.0cm}
\frac{{\rm d}R}{{\rm d} t} = u
\end{eqnarray}
One can solve these equations numerically for known values of $R_0$,
$M_0$, $u_0$, $\rho_{ISM}$, $L_{SC}$ and $V_{\infty}$.

For the particular case when supernovae and stellar winds deposit
all momentum instantaneously and the pressure in the interstellar
medium is zero, $P_{ISM} = 0$, equations (\ref{eq1.a}) and
(\ref{eq1.b}) become:
\begin{eqnarray}
      \label{eq4.a}
      & & \hspace{-1.0cm}
M(R) = M_0 + \frac{4 \pi}{3} \rho_{ISM} (R^3 - R^3_0)  ,
      \\[0.2cm]
      \label{eq4.b}
      & & \hspace{-1.0cm}
M(R) u(t) = M_0 u_0 .
\end{eqnarray}
Furthermore, neglecting the mass of the star forming cloud and
assuming
\begin{equation}
      \label{eq5}
M_0 = \frac{4 \pi}{3} \rho_{ISM} R^3_0 ,
\end{equation}
the solutions are reduced to the set of the main equations
(equations 1 and 2) used by Rela\~no et al. (2007):
\begin{eqnarray}
      \label{eq6.a}
      & & \hspace{-1.0cm}
R(t) = R_0 \left[1 + \frac{4 u_0 (t - t_0)}{R_0} \right]^{1/4} ,
      \\[0.2cm]
      \label{eq6.b}
      & & \hspace{-1.0cm}
u(t) = \frac{3 M_0 u_0}{4 \pi \rho_{ISM} R^3} .
\end{eqnarray}
Thus, their main equations represent an oversimplified 1D model that
neglects the effects of the ambient pressure, the mass of the star
forming cloud and the continuous mechanical energy injection.

To illustrate the differences in the solutions of equations
\ref{eq3.a} and \ref{eq3.b} and equations \ref{eq6.a} and
\ref{eq6.b}, we assume an interstellar gas density
\begin{equation}
      \label{a1}
\rho_{ISM} = \frac{3 M_{HI}}{4 \pi R^3_{HI}} ,
\end{equation}
where $M_{HI}$ and $R_{HI}$ are the mass and radius of an evolved HI
shell (for the comparison we use the particular case of GSH
285-02+86, whose radius and mass are $R_{HI} = 385$~pc and $M_{HI} =
44\times10^5$ M$_{\odot}$, respectively). The number density of the
interstellar gas then is $n_{ISM} = 0.75$~cm$^{-3}$. We further
assume that the initial radius of the shell is $R_0 = 104$~pc and
obtain the initial mass of the shell, $M_0$, from equation
(\ref{eq5}). Then we use the observed velocity of the progenitor
shell (64.7 km~s$^{-1}$) as the initial value for the solution of
equations \ref{eq3.a} and \ref{eq3.b}. For the oversimplified case
of equations \ref{eq6.a} and \ref{eq6.b} we use energy conservation,
and the initial velocity in this case is
\begin{equation}
      \label{a2}
u_0 = \frac{2 E_{kin}}{M_0} ,
\end{equation}
where the kinetic energy of the progenitor shell, $E_{kin} =
36.1\times10^{52}$~erg s$^{-1}$, has been derived from the
Starburst99 synthetic model (Leitherer et al. 1999). Note that one
cannot use identical initial conditions (the same initial velocity)
in both approaches because in one case the energy and momentum are
deposited instantaneously and, in the other, they are supplied
continuously and grow with time during the shell evolution.

Figure 2 presents the expansion velocities predicted by both sets of
equations. Certainly, the initial velocity in the oversimplified
model is much larger than that of equations \ref{eq3.a} and
\ref{eq3.b}, but it drops faster. When the radius of the shell
becomes about three or four times that of the initial value, the
difference between the two calculations becomes small. Later on, the
effect of the ambient pressure becomes important, but the shell
continues to expand in the oversimplified model.

In the next section we drop all simplifications associated with the
analytic solutions and solve equations (\ref{eq3.a}) and
(\ref{eq3.b}) numerically.

\section{The expansion of the momentum-dominated shell into the
         homogeneous ISM}

\subsection{Initial conditions}

We start the integrations at the initial time, $t_0 = 10^6$ yr, and
with the initial mass, $M_0$. We adopt initial radii, $R_0$, and
expansion velocities, $u_0$, derived from the H$_{\alpha}$
observations of the ionized shells. $R_0$ is approximately 0.3 times
the radius of the HII region, and $u_0$ is the observed velocity of
the H$_{\alpha}$ shell. For example, in the case of NGC 1530-8, $R_0
= 104$~pc and $u_0 = 64.7$~km~s$^{-1}$, respectively. In order to
calculate the initial mass of the shell, we substitute into equation
(\ref{eq5}) the value of the initial radius and the average density
 0.1~cm$^{-3}$ assumed by Rela\~no et al. We also
assume that the embedded cluster continuously expels the gas
released by SNe and stellar winds, whose momentum supports the
expansion of the outer shell, during $\sim 40$~Myr, the
characteristic life-time of a $~8$\Msol \, star - the lowest mass
star which will eventually explode as a supernova. Also, the
pressure in the surrounding medium is explicitly included here.

We calculate the density of the ISM using the masses and radii of
the HI shells and their progenitors:
\begin{equation}
      \label{eq7}
\rho_{ISM} = \frac{M_{HI} - M_0}{\frac{4 \pi}{3} (R^3_{HI} - R^3_0)}
\end{equation}
Thus, we derive the density of the ISM for each couple of HI and HII
shells from their observed parameters. For instance, in the case of
GSH 285-02+86, GSH 304--00-12 and GSH 305+01-24 from the list of
McClure-Griffiths et al. (2002), the interstellar gas number density
would be 0.75, 1.38 and 0.62 cm$^{-3}$, respectively, if one uses as
the progenitor the HII shell N8 found by Rela\~no et al. (2007) in
the spiral galaxy NGC~1530.

The average mechanical luminosity of the embedded cluster has been
calculated from the Starburst99 synthetic model with an
instantaneous burst of star formation (Leitherer et al. 1999). The
average mechanical luminosity of the cluster then is:
\begin{equation}
      \label{eq8}
L_{SC} = E_{SW+SN} / \tau ,
\end{equation}
where we use $\tau = 10$ Myr. The mechanical luminosities of the
embedded clusters, as well as the initial masses, radii and
velocities for all shells are listed in Table~1.

The temperature of the interstellar medium is assumed to be $T_{ISM}
= 6000$K, which is a typical value in the warm neutral component of
the ISM (Brinks, 1990). The thermal pressure in the ambient medium
then is $P_{ISM} = k~n_{ISM}~T_{ISM}$, where $k$ is the Boltzmann's
constant and $n_{ISM} = \rho_{ISM} / m_{H}$ is the interstellar gas
number density obtained from equation (\ref{eq7}).

The last input parameter for our model, the terminal speed of the
star cluster wind $V_{\infty}$, is determined by the energy and mass
deposition rates, and is close to the terminal velocity of
individual stellar winds (Raga et al. 2001, Stevens \& Hartwell
2003). We assume for our calculations that the star cluster wind
terminal speed is constant and falls in the range $1500-3000$ km
s$^{-1}$ (e.g., Leitherer at al. 1999). Note that this parameter
defines the amount of momentum deposited by the cluster and
therefore affects the dynamics of the momentum-dominated shell.

\subsection{Results and discussion}

We calculate the evolutionary tracks for the HII shells associated
with the list of large HII regions studied by Rela\~no et al. (2007), and
compare them with the observed parameters of the HI shells. For the
comparison we have chosen three HI shells from the list of
McClure-Griffiths et al. (2002) whose morphologies are close to the
spherical shape: GSH 285-02+86, GSH 304-00-12 and GSH 305+01-24.
They are representative of high, intermediate and low mass objects,
respectively.

The results of the calculations are presented in Figure 3 and in
Table 2. The top left panel in Figure 2 compares the results of the
calculations with the parameters of low mass neutral shells. It
shows that most of the HII shells detected in the giant HII regions
certainly can evolve into objects whose parameters are identical to
those of low mass HI shells. For example, in the case of GSH
305+01-24, nine out of the twelve HII shells (see Table 2) can
easily reach the observed size of the HI shell having a proper mass
($3.9\times10^5$ M$_{\odot}$) and an expansion velocity which is similar 
or even higher than the observed velocity of GSH 305+01-24. This implies 
that the HII shells found by Rela\~no et al. (2007) in giant HII regions 
can be progenitors of such low mass HI shells.

The top right panel in Figure 3 compares the evolutionary tracks of
the progenitor shells with the observed parameters of the
intermediate mass ($1.9\times10^6$ M$_{\odot}$) HI object GSC
304-00-12. In this case only the four most energetic HII shells (NGC
1530-8, NGC 1530-22, NGC 3359-6 and NGC 6951-2) can eventually reach
the required radius sweeping enough interstellar mass (see Table 2).
However only two of them (NGC 1530-8 and NGC 6951-2) present the
expansion velocities that are similar to the observed one. The
expansion velocities of the rest fall well below the observed value.
Thus GSC 304-00-12 represents a limit case that separates the low
mass HI objects driven by multiple SNe explosions from the largest
ones whose parameters are not consistent with the multiple SNe
hypothesis. The latter case is illustrated by the bottom panel in
Figure 3.

The bottom panel compares the evolutionary tracks of Rela\~no's
shells with parameters of the very massive shell GSH 285-02+86,
whose mass is $4.4 \times 10^6$ M$_{\odot}$ (McClure-Griffiths et
al. 2002). Here we also present (dashed line) the expansion velocity
predicted by the analytic model for NGC 1530-8, the most energetic
shell from the list of Rela\~no et al. (2005). The plot clearly
demonstrates that the analytic model leads to overestimated
expansion velocities. This is because the analytic calculations
adopt a low value for the interstellar gas density which is not
consistent with the masses and radii of the most massive neutral
shells (in the analytic approach the mass of the shell is only $0.6
\times 10^6$ \Msol \, when its radius reaches the observed value of
385~pc, whereas in our calculations it is $4.4 \times 10^6$ \Msol).

The bottom panel in Figure 3 shows that HII shells found by Rela\~no
et al. (2005) inside giant HII regions can never evolve into the
largest and most massive neutral hydrogen supershells. In this case
the energy and momentum supplied by the central cluster are not
sufficient to make the evolution from the original shell finally fit
all the observed parameters of the HI supershell: its mass, radius
and expansion velocity. In this case only the three most energetic
progenitor shells (NGC 1530-8, NGC 3359-6 and NGC 6951-2) can reach
a size and accumulate a mass that are comparable to those of GSH
285-02+86. However, in all these cases the model predicted expansion
velocities are much smaller than that measured by McClure-Griffiths
et al. (2002) for GSH 285-02+86. This implies that, contrary to the
statement made by Rela\~no et al. (2007), the young ionized shells
found in giant HII regions cannot evolve into the largest HI
supershells (i.e., eventually fit simultaneously the mass, radius
and expansion velocity of the HI object) unless some additional
physical mechanism, other than the multiple supernovae explosions,
contributes to the formation of the largest shells at the later
stages of their evolution.

Figure 3 presents the results of the calculations under the
assumption that the star cluster terminal speed is $V_{\infty} =
1500$~km s$^{-1}$. In order to learn how this parameter affects our
results we have provided a set of numerical calculations with
initial conditions which are identical to those used in the case of
GSH 285-02+86 (Figure 3, bottom panel) but with a star cluster wind
terminal speed twice as large as the previous one: $V_{\infty} =
3000$~km~s$^{-1}$. The results of these calculations are presented
in Figure 4 which demonstrates that the model predicted expansion
velocity becomes smaller when the star cluster wind terminal speed
grows up. This implies that one cannot avoid the discrepancy between
the predicted and the observed velocities by varying the star
cluster wind terminal speed within a reasonable velocity interval.

\section{Conclusions}

Here we have critically examined the evolution of the ionized shells
found inside giant HII regions, as predicted by the multiple SNe
model in one dimension. We have used the observed parameters of the
HII shells as initial conditions for our numerical model and
compared the results of the calculations with three representative
cases of low, intermediate and high mass HI objects from the list of
McClure-Griffiths et al. (2002).

We have found that the ionized shells observed within giant HII
regions cannot evolve into the largest neutral hydrogen supershells
if the multiple supernovae explosions of massive stars is the only
driving mechanism. Some additional physical mechanism must
contribute to the formation of the largest shells for the model to
be in agreement with the observed parameters of the most massive
neutral hydrogen supershells detected in our and other galaxies.

Note that spherically-symmetric models must be taken with care when
compared with objects whose radii are comparable with several
characteristic scale heights in the ISM density distribution. If
that is the case, the expansion velocities at the top and at the
waist of the shell are different and the shell acquires a distorted
hour-glass form. The majority of the swept-up interstellar matter is
concentrated in a thin layer nearby the plane of the host galaxy. 3D
calculations are then required in order to investigate the shell's
morphology and kinematics in the differentially rotating galactic
disk (see, for example, Silich et al. 1996).

\begin{acknowledgements}
We thank an anonymous referee for central comments and suggestions
which helped us to clarify our formulations and improve the quality
of the paper. We also appreciate usuful comments from our Korean and
Mexican colleagues during the IV Korea-Mexico workshop in Daejon.
This study has been supported by Conacyt (M\'exico) grant 47534-F.
\end{acknowledgements}

\clearpage

\begin{figure}
\includegraphics[angle=0,scale=0.60]{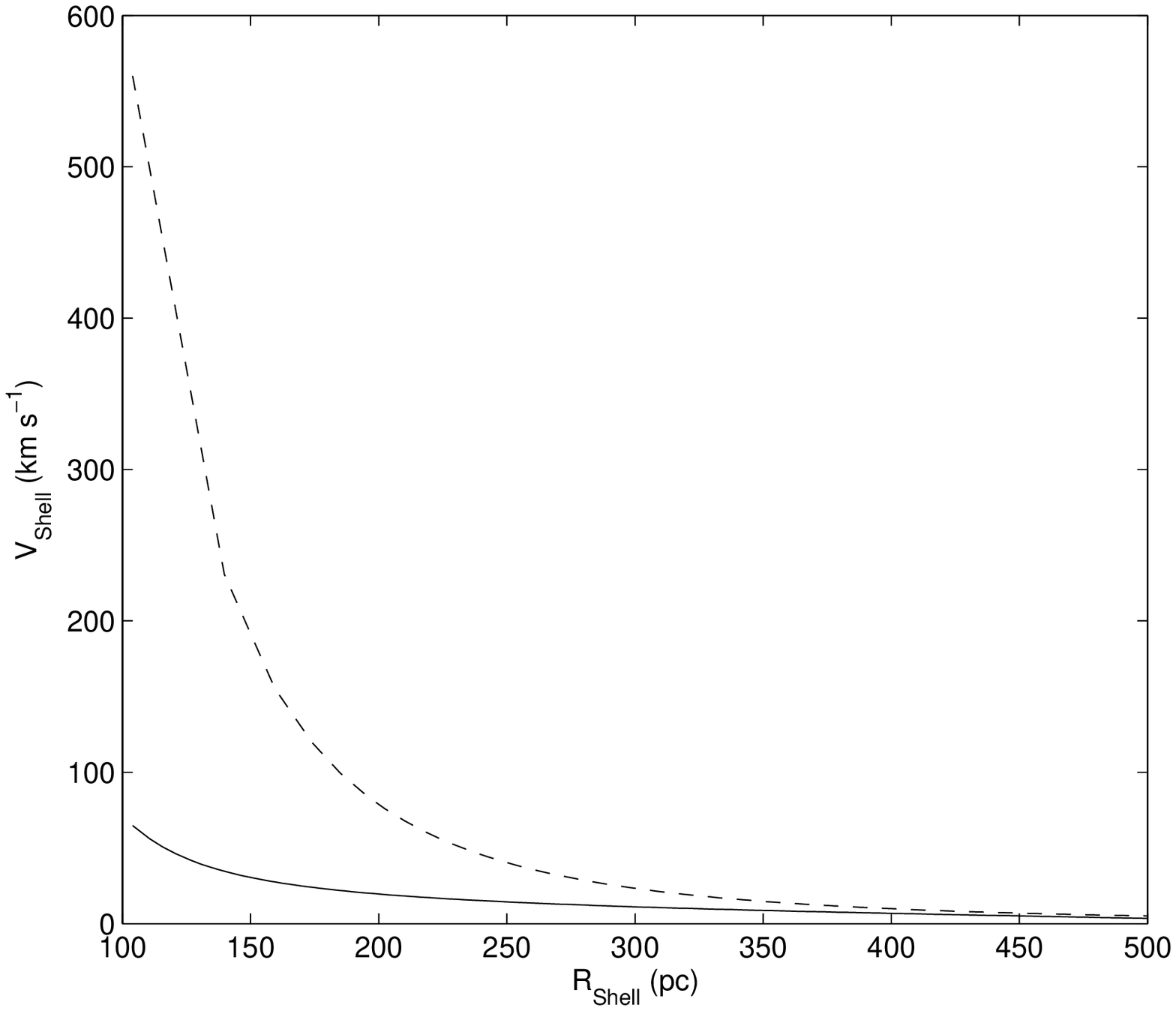}
\caption{The comparison of analytic results with numerical models (the
example corresponds to the HI shell GSH 285-02+86 and the HII progenitor 
shell NGC 1530-8). The solid line represents the numeric
solution of equations \ref{eq3.a} and \ref{eq3.b}. The dashed line
is the analytic solution of equations \ref{eq6.a} and \ref{eq6.b}.
The initial radii and masses of the shells are identical in both
calculations, however the initial velocities are different. This is
because the analytic formulation is based on the assumption that all
energy and momentum are deposited instantaneously at the beginning
of the momentum dominated stage, whereas the numerical model assumes
that the energy and momentum are supplied continuously until the
last supernova explosion.} \label{fig2}
\end{figure}

\begin{figure}
\vspace{14.0cm} \includegraphics{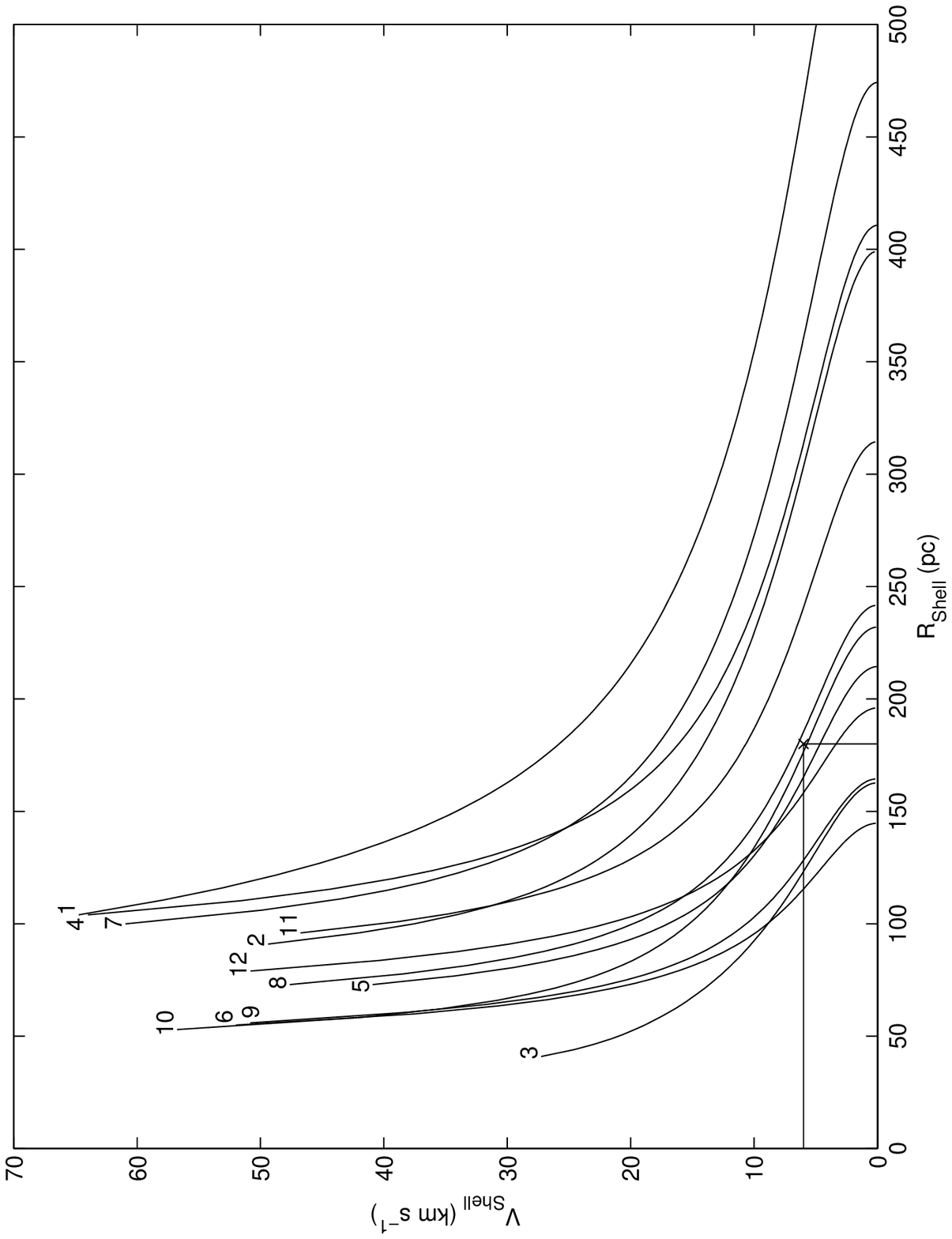}
\includegraphics{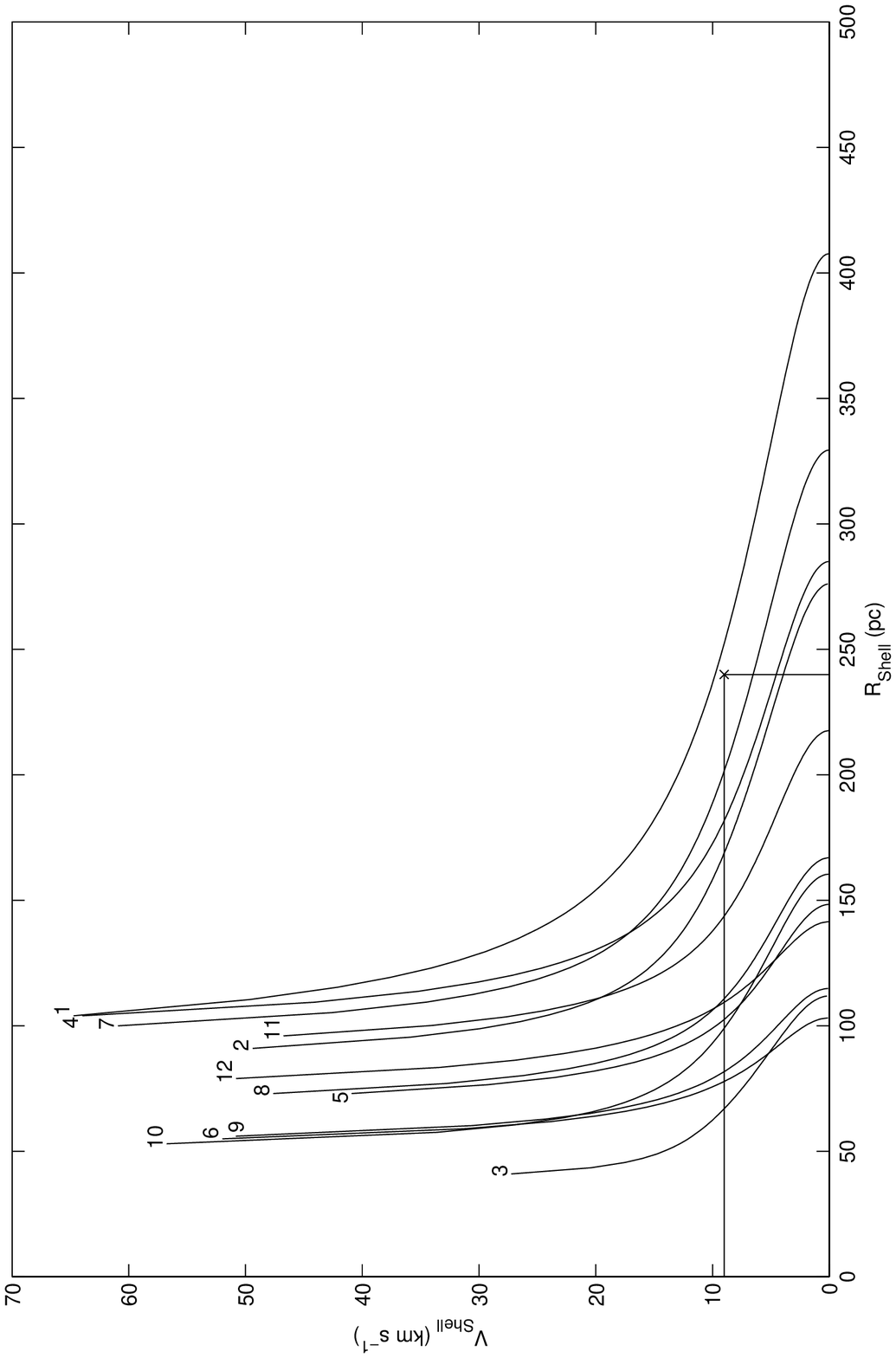}
\includegraphics{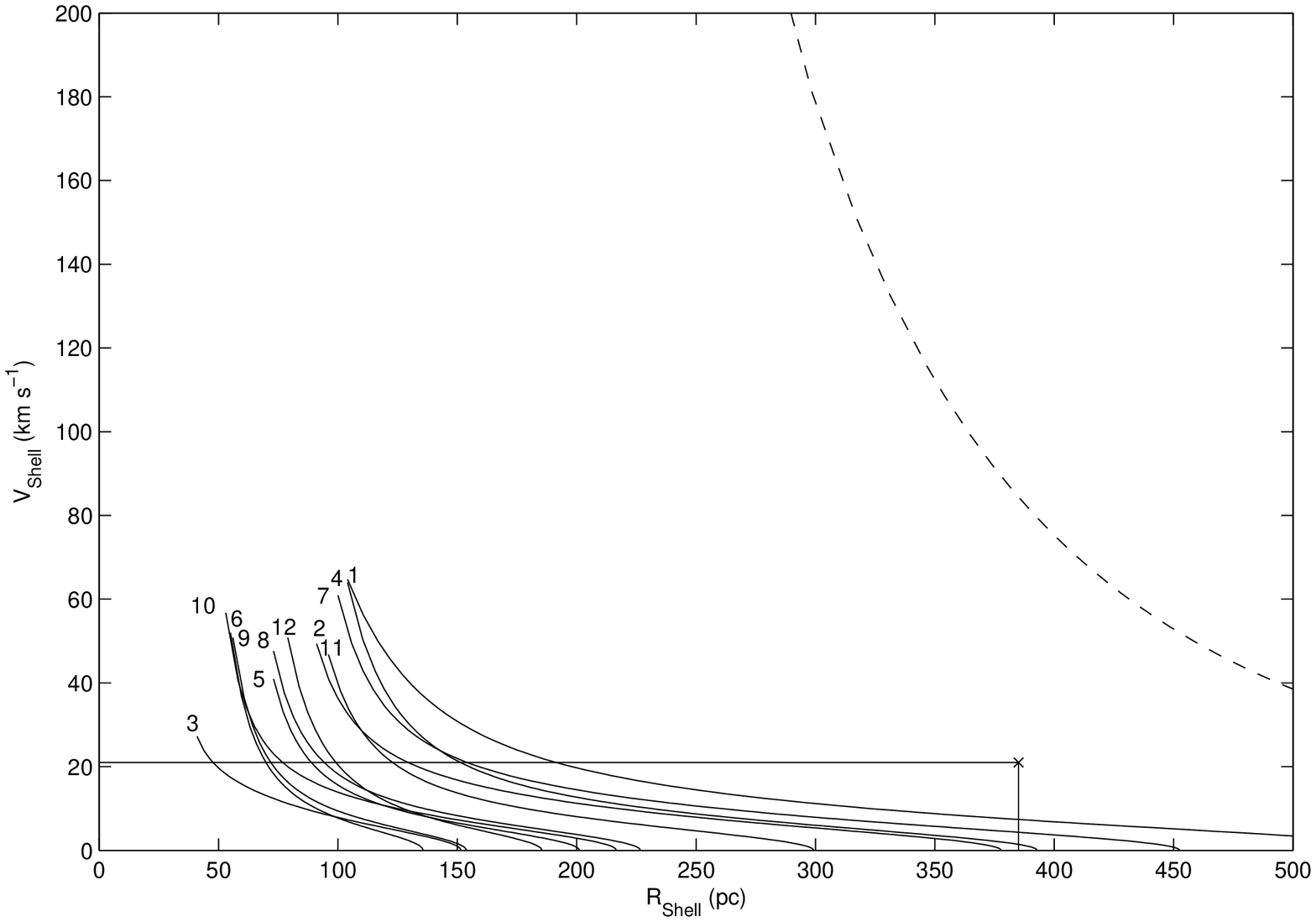}
\caption{The comparison of the predicted expansion velocities with
the observed values. The results of the calculations are compared
with parameters of low, intermediate and high mass objects from the
list of McClure-Griffiths et al. (2002). Different lines correspond
to different initial conditions (associated with 12 HII progenitor
shells from the list of  Rela\~no et al. 2007). The lines are
labeled with the numbers that identify the progenitor shells in
Table 1. The observed velocities and radii of the HI shells are
marked by the horizontal and vertical lines, respectively. The top
left panel compares the model predicted radii and velocities with
that of the HI supershell GSH 305+01-24 from the list of
McClure-Griffiths et al. (2002). Similarly, the top right and bottom
panels present the results for the more massive supershells GSH
304-00-12 and GSH 285-02+86 (the dashed line in the latter is
Rela\~no's et al. analytic solution for NGC 1530-8). A
1500~km~s$^{-1}$ star cluster wind terminal speed was adopted for
all calculations.} \label{fig3}
\end{figure}


\begin{figure}
\includegraphics[angle=270,scale=0.60]{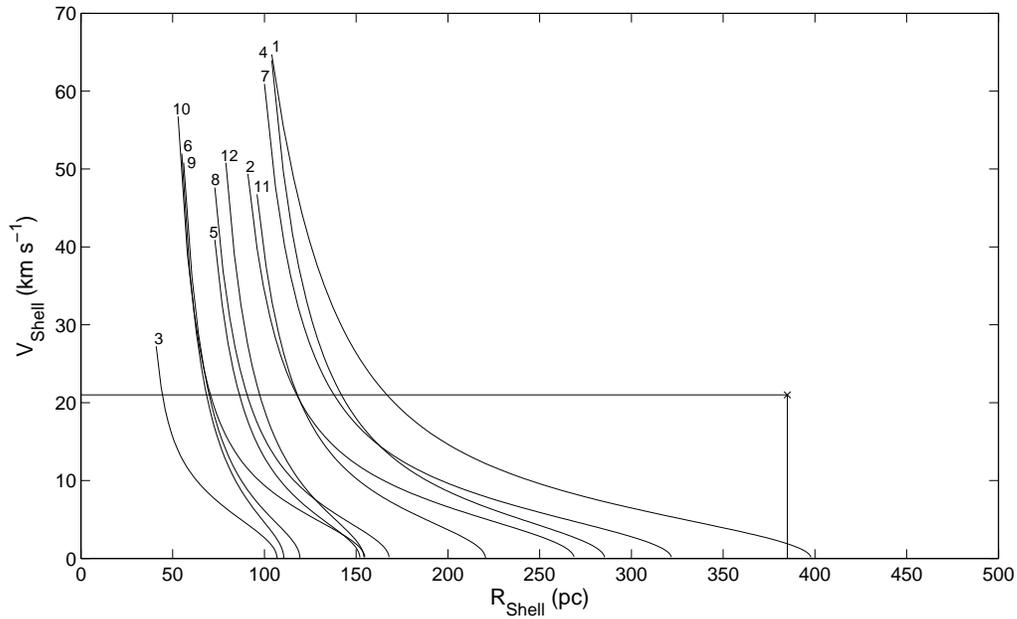}
\caption{The impact of the star cluster wind terminal speed on the
shell evolution. The calculations were performed for the HI shell
GSH 285-02+86, with a star cluster wind terminal speed $V_{\infty} =
3000$ km s$^{-1}$. This can be compared with that presented in
Figure 3 (bottom panel) where a value of $V_{\infty} = 1500$ km
s$^{-1}$ for the wind terminal speed has been adopted.} \label{fig4}
\end{figure}

\clearpage

\begin{deluxetable}{rlcccc}
\tablecolumns{6} \tablewidth{0pt} \tabletypesize{\scriptsize}
\tablecaption{Initial conditions\label{tbl-1}}

\tablehead{\colhead{ID} & \colhead{Name} & \colhead{$M_0$} &
\colhead{$R_0$} & \colhead{$u_0$} & \colhead{$L$}\\
\colhead{} & \colhead{} & \colhead{$10^4 M_{\odot}$} &
\colhead{$pc$} & \colhead{$km~s^{-1}$} & \colhead{$10^{38}
erg~s^{-1}$}}

\startdata

$1$  & NGC 1530--8   & $8.9$ & $104$ & $64.72$ & $11.4$ \\
$2$  & NGC 1530--22  & $3.7$ & $91$  & $49.38$ & $5.3$  \\
$3$  & NGC 1530--92  & $0.3$ & $41$  & $27.24$ & $0.9$  \\
$4$  & NGC 3359--6   & $5.5$ & $104$ & $63.97$ & $5.7$  \\
$5$  & NGC 3359--42  & $1.9$ & $73$  & $40.91$ & $1.5$  \\
$6$  & NGC 3359--92  & $0.8$ & $55$  & $51.98$ & $0.6$  \\
$7$  & NGC 6951--2   & $4.8$ & $100$ & $60.93$ & $7.6$  \\
$8$  & NGC 6951--18  & $1.9$ & $73$  & $47.62$ & $1.9$  \\
$9$  & NGC 6951--41  & $0.8$ & $56$  & $50.82$ & $0.9$  \\
$10$ & NGC 5194--312 & $0.7$ & $53$  & $56.75$ & $1.7$  \\
$11$ & NGC 5194--403 & $4.2$ & $96$  & $46.75$ & $3.3$  \\
$12$ & NGC 5194--416 & $2.3$ & $79$  & $50.80$ & $1.2$  \\

\enddata
\end{deluxetable}

\clearpage
\begin{sidewaystable}
{\scriptsize \caption{Model predictions\label{tbl-2}}
\begin{tabular}{rl|ccc|ccc|ccc}

\multicolumn{2}{c}{} & \multicolumn{3}{|c}{GSH 285--02+86} &
\multicolumn{3}{|c}{GSH 304--00-12} & \multicolumn{3}{|c}{GSH 305+01-24}\\
\tableline \tableline \\
ID & Name & $M$ & $R$ & $V$ & $M$ & $R$ & $V$ & $M$ & $R$ & $V$\\
 &  & $10^5 M_{\odot}$ & $pc$ & $km s^{-1}$ & $10^5 M_{\odot}$ & $pc$ & $km s^{-1}$ & $10^5 M_{\odot}$ & $pc$ & $km
 s^{-1}$\\

\cutinhead{Observed parameters}

& & $44$ & $375-395$ & $21$ & $19$ & $200-280$ & $9$ & $3.9$ & $140-220$ & $6$ \\

\cutinhead{Predicted parameters}

$1$  & NGC 1530--8   & $44$  & $385$ & $7.4$       & $19$  & $240$ & $9.7$     & $3.9$ & $179$ & $26.1$ \\
$2$  & NGC 1530--22  & $42$  & $378$ & $0.1$       & $19$  & $240$ & $3.9$     & $3.9$ & $179$ & $14.2$ \\
$3$  & NGC 1530--92  & $27$  & $152$ & $0.0$       & $19$  & $112$ & $0.2$     & $2.9$ & $163$ & $0.2$  \\
$4$  & NGC 3359--6   & $44$  & $385$ & $1.4$       & $19$  & $240$ & $4.5$     & $3.9$ & $180$ & $16.1$ \\
$5$  & NGC 3359--42  & $6.2$ & $201$ & $0.1$       & $4.3$ & $148$ & $0.0$     & $3.9$ & $180$ & $4.6$  \\
$6$  & NGC 3359--92  & $1.9$ & $136$ & $0.0$       & $14$  & $103$ & $0.1$     & $2.0$ & $145$ & $0.1$  \\
$7$  & NGC 6951--2   & $44$  & $385$ & $4.3$       & $19$  & $240$ & $6.5$     & $3.9$ & $180$ & $17.8$ \\
$8$  & NGC 6951--18  & $9.0$ & $227$ & $0.0$       & $6.2$ & $167$ & $0.1$     & $3.9$ & $180$ & $6.5$  \\
$9$  & NGC 6951--41  & $2.8$ & $154$ & $0.0$       & $1.9$ & $115$ & $0.1$     & $3.0$ & $165$ & $0.2$  \\
$10$ & NGC 5194--312 & $7.9$ & $217$ & $0.1$       & $5.6$ & $160$ & $0.1$     & $3.9$ & $180$ & $5.7$  \\
$11$ & NGC 5194--403 & $21$  & $299$ & $0.1$       & $14$  & $218$ & $0.0$     & $3.9$ & $180$ & $10.7$ \\
$12$ & NGC 5194--416 & $4.8$ & $185$ & $0.1$       & $3.5$ & $142$ & $0.0$     & $3.9$ & $180$ & $3.4$  \\

\end{tabular} }
\end{sidewaystable}


\begin{thebibliography}

\bibitem{1} Bisnovatyi-Kogan, G. S. \& Silich, S. A. 1995, Rev. Mod.
            Phys. 67, 661

\bibitem{2} Brinks, E., 1990, ASSL, 161, 39

\bibitem{3} Brinks, E. \& Bajaja, E. 1986, A\&A, 169, 14

\bibitem{4} Bruhweiler, F.C., Gull, T.R., Kafatos, M. \& Sofia, S.
            1980, ApJL, 238, 27

\bibitem{5} Castor, J., McCray, R. \& Weaver, R., 1975, ApJ, 200, 107

\bibitem{6} Chu, Y.-H. \& Mac Low, M.-M. 1990, ApJ, 365, 510

\bibitem{7} Comeron, F. \& Torra, J. 1992, A\&A, 349, 41

\bibitem{8} Crosthwaite L.P., Turner, J.L. \& Ho, P.T.P. 2000, AJ, 119, 1720

\bibitem{9} Dib, S. \& Burkert, A. 2004, Ap\&SS, 292, 135

\bibitem{10} Dyson, J.E. 1980, Physics of the Interstellar Medium,
             NY, John Wiley \& Sons, 145

\bibitem{11} Efremov, Y. N., Ehlerov\'a, S. \& Palou\v s, J. 1999,
            ApJ, 350, 457

\bibitem{12}  Ehlerov\'a, S. \& Palou\v s, J. 2005, A\&A, 437, 101

\bibitem{13} Elmegreen, B. G. \& Chiang, W.-H. 1982, ApJ, 253, 666

\bibitem{14} Gil de Paz, A., Silich, S.A., Madore, B.,F., S\'anchez Contreras,
C., Zamorano, J. \& Gallego, J. 2002, ApJ, 573, 101

\bibitem{15} Hatzidimitriou, D., Stanimirovic, S., Maragoudaki, F.,
            Stavely-Smith, L., Dapergolas, A. \& Bratsolis, E.
            2005, MNRAS, 360, 1171

\bibitem{16} Heiles, C. 1980, ApJ, 235, 833

\bibitem{17} Heiles, C. 1984, ApJS, 55, 585

\bibitem{18} Kim, S., Dopita, M. A., Stavelet-Smith, L. \&
             Bessel, M. 1999, A\&A, 350, 230

\bibitem{19} Koo, B.-C. \& McKee, C.F., 1992, ApJ, 388, 93

\bibitem{20} Leitherer, C., et al. 1999, ApJS, 123, 3

\bibitem{21} Lozinskaya, T.A. 1992, Supernova and Stellar Wind in the
             Interstellar Medium, AIP, NY, 223p.

\bibitem{22} Lozinskaya, T., Moiseev, A., Podorvanyuk, N., 2003,
             Astronomy Letters, 29, 77, (Astro-ph/0301214)

\bibitem{23} Mac Low, M.-M. \&  McCray, R. 1988, ApJ, 324, 776

\bibitem{24} Mashchenko, S.Y., Thilker, D.A. \& Braun, R. 1999, A\&A, 343, 352


\bibitem{26} McClure-Griffiths, N.M., Dickey, J.M., Gaensler, B.M. \&
            Green, A.J. 2002, AJ, 578, 176

\bibitem{27} McCray, R. \& Kafatos, M. 1987, ApJ, 317, 190

\bibitem{28} Naz\'e, Y., Chu, Y.-H., Points, S.D., Danforth, C.W.;
             Rosado, M. \& Chen, C.-H.R.


\bibitem{30} Oey, M.S. \& Garc\'\i a-Segura, G., 2004, ApJ, 613, 302

\bibitem{31} Oey, M.S. \& Massey, P. 1995, ApJ, 452, 210

\bibitem{32} Palous J., Franco, J. \& Tenorio-Tagle, G. 1990,
             Astron. Astrophys. 227, 175

\bibitem{33} Puche, D., Westpfahl, D., Brinks, E. \& Roy, J-R. 1992, AJ, 103,
             1841

\bibitem{34} Raga, A.C., Vel\'azquez, P.F., Cant\'o, J., Masciadri, E.,
             Rodr\'\i guez, L.F., 2001, ApJ, 559, 33

\bibitem{35} Rela\~no, M. \& Beckman, J.E. 2005 A\&A, 430, 911

\bibitem{36} Rela\~no, M., Beckman, J.E., Daigle, O., Carignan, C. 2007,
             A\&A, 467, 1117

\bibitem{37} Rhode, K. L., Salzer, J. J., Westpfahl, D. \& Radice, L. A.
             1999, AJ, 118, 323

\bibitem{38} Rozas, M., Beckman, J.E. \& Knapen, J-H. 1996, A\&A, 307, 735

\bibitem{39} Silich, S.A. 1992, Ap\&SS 195, 317.

\bibitem{40} Silich, S.A., Franco, J.,  Palou\v{s} \&  Tenorio-Tagle, G.
             1996, ApJ, 468, 722



\bibitem{41} Silich, S.A., Tenorio-Tagle G.,
             Mu\~noz-Tu\~non, C., Cairos L.-M. \&
             Gil de Paz A., 2002, in ASP Conf. Ser. 282 ``Galaxies: The Third
             Dimension'' Edts. M. Rosado, L. Binette \& L. Arias,
             p58

\bibitem{42}  Silich, S., Lozinskaya, T., Moisseev, A., Podorvanuk, N.,
       Rosado, M., Borissova, J. \& Valdez-Guti\'errez, M.
       2006, Astron. Astrophys., 448, 123

\bibitem{43} Smith, D.A. \& Wang, Q.D., 2004, ApJ, 611, 881

\bibitem{44} Stevens, I.R. \& Hartwell, J.M., 2003, MNRAS, 339, 280

\bibitem{45} Tenorio-Tagle, G. 1981, Astron. Astrophys. 94, 338

\bibitem{46} Tenorio-Tagle, G. \& Bodenheimer, P., 1988, ARA\&A, 26, 145

\bibitem{47} Valdez-Guti\'errez M., Rosado M., Georgiev L.,  Borissova J.,
             Kurtev R. 2001, Astron. Astrophys., 366, 35.

\bibitem{48} Wada, K. \& Norman, C.A. 1999, ApJ, 516, L13

\bibitem{49} Wada, K., Spaans, M. \& Kim, S., 2000, ApJ, 540, 797

\bibitem{50} Weaver, R., McCray, R., Castor, J., Shapiro, P., Moore,
R., 1977, ApJ, 218, 377

\end{thebibliography}
\end{document}